\documentclass[12pt]{article}  
\usepackage[dvips]{graphicx}
\usepackage{wrapfig}
\usepackage{bm}
\def\bar{\overline}
\def\e{\epsilon}
\def\ord{{\mathcal O}}
\begin{document}    
\baselineskip=24.5pt    
\setcounter{page}{1}         
\topskip 0 cm    
\vspace{1 cm}    
\centerline{\Large\bf  Examining the Geometrical Model }
\centerline{\Large\bf  with Inverted Mass Hierarchy for Neutrinos}
\vskip 1 cm
\centerline{\large Mizue Honda$^1$ and Morimitsu Tanimoto$^2$}
\vskip 5mm
\centerline{$^1$  Graduate School of Science and Technology, 
Niigata University, Niigata 950-2181, Japan}
\centerline{$^2$ Department of Physics, Niigata University, Niigata 950-2181,
Japan}

 
\vskip 4 cm
\noindent
{\large\bf Abstract}

The comprehensive analyses are presented in the model with 
the inverted mass hierarchy for neutrinos, which follows from 
a geometrical structure of a (1+5) dimensional space-time where two extra
dimensions are compactified on the ${\bf T^2/Z_3}$ orbifold.
The model gives two large lepton flavor mixings due to the $S_3$ structure 
in the (1+5) dimensional space-time.
It also predicts the lightest neutrino mass as 
$m_3=(1$-$50)\times 10^{-5}$ eV and the effective neutrino mass  
responsible for  neutrinoless double beta decays as
$\langle m\rangle_{ee}\simeq 50$ meV.
The low energy $CP$ violation, $J_{CP}$ could be $0.02$.
On the other hand, the observed baryon asymmetry in the present universe is
produced by the non-thermal leptogenesis, which works 
even at the reheating temperature  $T_R=10^{4}$-$10^{6}\ {\rm GeV}$.
The correlation between the baryon asymmetry and the low energy $CP$ 
violation is examined in this model.  

\newpage 

\section{Introduction}
It is the important task to find  an origin of the observed hierarchies 
 in masses and flavor  mixings for quarks and leptons.
In particular, the geometrical aspect 
provides a progressive study for masses and mixings of quarks and leptons.
  The non-Abelian discrete flavor symmetry is also realized 
in the simple geometrical understanding of  superstring theory \cite{string}.
Recently, a higher-dimensional model of neutrinos \cite{TY} was proposed   
 based on the orbifold  model \cite{Geometry}.
In this  model,  extra two dimensions  in  a (1+5) dimensional space-time  
are compactified on the ${\bf T^2/Z_3}$ orbifold, which has three equivalent
 fixed points.  The quarks and leptons are supposed to belong to
 ${\bf 5}^*$ and  $\bf 10$ in the SU(5) grand unification model.
A ${\bf 5}^*$ and a right-handed neutrino $N$ in each 
family are localized on each of the  three fixed points of the orbifold, 
while three ${\bf 10}$'s live in  the bulk. 
Assuming the discrete flavor symmetry $S_3$, 
the successful democratic mass matrices for quarks and leptons are obtained,
 provided that 
the three ${\bf 10}$'s and Higgs doublets are  distributed 
homogeneously in the bulk. 
 On the other hand, the Higgs $\phi$ for the B-L breaking is localized 
 on one of the three fixed points. Therefore,  one of $N$ acquires 
a superheavy Majorana mass generating an inverted hierarchy 
in the neutrino mass spectrum. 

The previous work  \cite{TY} has shown that 
 the  model is consistent with  observations 
on the neutrino masses and mixings. 
In particular,  the leptogenesis  works well 
 with the reheating temperature  $T_R=10^{7}$-$ 10^{10}\  {\rm GeV}$.
However, a CP violating phase is only taken account for simplicity
 in the previous work.
In this paper, we present comprehensive numerical analyses 
 including all possible  CP violating phases
 to test the model  in future  experiments.
Moreover, we scan  wider regions of   parameters to be consistent with
all data of  the neutrino masses and mixings.
We have found that  the leptogenesis  works 
at  the lower reheating temperature  $T_R\simeq 10^{4}$-$ 10^{6}\ {\rm GeV}$.

 In section 2, we summarize the model for   the neutrino and 
charged lepton mass matrices,
where the inverted neutrino mass spectrum is naturally obtained.
In section 3,
 numerical analyses are given for the neutrino masses and mixings.
In  section 4, the non-thermal leptogenesis is numerically studied.
 The correlation between the leptogenesis and the low energy $CP$
violation is also examined. 
The summary is devoted to section 5.
 
\section{Framework of the Model}

\begin{wrapfigure}{r}{6 cm}
\begin{center}
    \includegraphics[width=5cm]{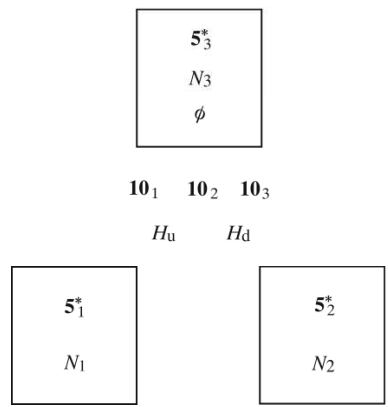}
\end{center}
\caption{The configuration of matters and Higgses at three fixed points
and in the  bulk.}
\end{wrapfigure}

Let us begin with presenting  the  model \cite{TY},
 which is based on works in \cite{Geometry,IIB}.
As shown in Fig. 1,
 a ${\bf 5_i}^*$ and a right-handed neutrino $N_i$  in each family are 
localized on one of the equivalent three fixed points of 
the ${\bf T^2/Z_3}$ orbifold 
while three ${\bf 10_i}$'s and Higgs doublets
 $H_u$ and $H_d$ live in the bulk. 
The $\phi$ for the B-L breaking is localized on the fixed point 
on which the ${\bf 5}^*_3$ and $N_3$ in the third family reside. 
Therefore, 
the $N_3$ in the third family has a very large Majorana mass compared
 with those of other $N$'s.

Since the three fixed points are  equivalent to one another,  we assume an
$S_3$ family symmetry acting on three ${\bf 5}^*$'s and on three $N$'s.
We also introduce breakings of the $S_3$ symmetry to obtain 
realistic mass matrices. We assume two sources of the breaking. One is a 
localization of the wave function of the $\phi$ field and 
the other is small distortions
of the wave functions of the three ${\bf 10}$'s and the doublet Higgses from 
homogeneous forms in the bulk.


\subsection{Neutrino  Mass Matrix}

The Dirac neutrino mass matrix 
is given by  the Yukawa coupling matrix  of $N{\bf 5}^* H_u$.
There are two independent matrices which are invariant of the $S_3$ symmetry \cite{FTY,FX}.
The $S_3$ invariant Dirac mass matrix is given by \cite{FTY}
\begin{equation}
h_\nu=h_0\left [\left (\matrix{1& 0& 0\cr 0 &1 &0\cr 0 & 0 &1 \cr      } \right )
+\left ( \matrix{0& \bar\epsilon & \bar\epsilon \cr
                  \bar\epsilon  & 0 & \bar\epsilon\cr
                  \bar\epsilon  & \bar\epsilon  & 0 \cr
                                         } \right ) \right ].
\end{equation}
\noindent
The parameter $\bar\epsilon$ is suppressed by separation of
the three fixed points \cite{Ark}. 

We have assumed  a homogeneous distribution of the Higgs fields $H_u$
in the bulk. However, 
 a small distortion of the Higgs wave function may induce a 
violation of the $S_3$ symmetry. 
The breaking effects appear first in the diagonal elements 
of the above mass matrix and the effects in the $\bar\epsilon$ term may 
be negligible for our discussion.
Therefore, the neutrino Dirac mass matrix is given by 
\begin{equation}
h_\nu=h_0\left 
[\left (\matrix{1& 0& 0\cr 0 &1+\bar\delta_1 &0\cr 0 & 0 &1+\bar\delta_2 \cr      } \right )
+\left ( \matrix{0& \bar\epsilon & \bar\epsilon \cr
                  \bar\epsilon  & 0 & \bar\epsilon\cr
                  \bar\epsilon  & \bar\epsilon  & 0 \cr
                                         } \right ) \right ] 
=h_0\left ( \matrix{1& \bar\epsilon & \bar\epsilon \cr
                  \bar\epsilon  & 1+\bar\delta_1 & \bar\epsilon\cr
                  \bar\epsilon  & \bar\epsilon  & 1+\bar\delta_2 \cr
                                         } \right )\ .
\end{equation}
\noindent
Here, $\bar\delta_1$, $\bar\delta_2$ and $\bar\epsilon$ 
are parametrized as 
\begin{eqnarray}
 \bar\delta_1=\delta_1 e^{i\beta_1},\qquad
\bar\delta_2=\delta_2 e^{i\beta_2},\qquad {\bar\epsilon}=\epsilon e^{i\varphi} \ ,
\end{eqnarray}
\noindent where $\delta_1$, $\delta_2$ and $\epsilon$ are real parameters,
 and $\beta_1$, $\beta_2$ and $\varphi$ are  CP violating phases.

The Majorana mass matrix $M_R$ for the right-handed neutrino $N_i ~(i=1,2,3)$ 
is determined also
by localization properties of $N_i$ and $\phi$ fields.
Since the $\phi$  is assumed to reside on one of the three fixed points where the third family
$N_3$ is localized, the (3,3) element of the Majorana mass matrix dominates 
over other elements. 
Then, the Majorana mass matrix is given  
as follows:
\begin{equation}
M_R= M_0\left ( \matrix{\bar\epsilon_2& \bar\epsilon_3^2 &\bar \epsilon_1 \cr
                  \bar\epsilon_3^2  & \bar\epsilon_2 & \bar\epsilon_1\cr
                  \bar\epsilon_1  & \bar\epsilon_1  & 1\cr   } \right )\ ,
\label{rightmatrix}
\end{equation}
\noindent
where  $\bar\epsilon_1$,  $\bar\epsilon_2$ and $\bar\epsilon_3$
are three independent  suppression factors.  These are parametrized as 
\begin{equation}
\bar\epsilon_1=\epsilon_1 e^{i\alpha_1},\qquad 
\bar\epsilon_2=\epsilon_2 e^{i\alpha_2},\qquad 
\bar\epsilon_3=\epsilon_3 e^{i\alpha_3},
\end{equation}
\noindent
where   $\epsilon_1$, $\epsilon_2$ and $\epsilon_3$ are real parameters,
 and $\alpha_1$, $\alpha_2$ and $\alpha_3$ are  CP violating phases.
Notice that the suppression factors come from the separation of the distinct fixed points and hence the parameters $\epsilon_i~(i=1,2,3)$ are the same order of the magnitude of $\epsilon $.

 Through the  seesaw mechanism \cite{Seesaw, Seesaw2},
the  effective neutrino mass matrix is given, in the leading order
of $\bar\epsilon,\ \bar\epsilon_i$ and $\bar\delta_i$, by
\begin{equation}
 M_\nu =h_\nu^T M_R^{-1}h_\nu \simeq
\frac{h_0^2}{\bar\epsilon_2 M_0}
\left (\matrix{1-\frac{\bar\epsilon_1^2}{\bar\epsilon_2}& 
  2\bar\epsilon+\tilde \epsilon & \bar\epsilon-\bar\epsilon_1 \cr
2\bar\epsilon+\tilde \epsilon& 1+2\bar\delta_1-
\frac{\bar\epsilon_1^2}{\bar\epsilon_2} 
& \bar\epsilon-\bar\epsilon_1\cr
  \bar\epsilon-\bar\epsilon_1  & \bar\epsilon-\bar\epsilon_1 & \bar\epsilon_2 \cr          } \right )\ ,
\label{mass}
\end{equation}
\noindent 
where 
\begin{equation}
\tilde\epsilon =\frac{\bar\epsilon_1^2-\bar\epsilon_3^2}{\bar\epsilon_2}\ .
\end{equation}
Since the neutrino mass matrix is complex, we discuss  
 $M_{\nu}M_{\nu}^{\dagger}$ :
\begin{eqnarray}
 M_{\nu}M_{\nu}^{\dagger}
 \simeq 
 \frac{h_0^4}{\epsilon_2^2 M_0^2}
 \left(
         \matrix{
              1-\frac{2\epsilon_1^2\cos{(2\alpha_1-\alpha_2)}}{\epsilon_2} 
                          &  4\epsilon\cos{\varphi}+2\tilde{\epsilon}'  
                          &  \bar{\epsilon}^*-\bar{\epsilon}_1^* \cr 
                   4\epsilon\cos{\varphi}+2\tilde{\epsilon}' 
   &  1-\frac{2\epsilon_1^2\cos{(2\alpha_1-\alpha_2)}}{\epsilon_2}+4\delta_1
                         &  \bar{\epsilon}^*-\bar{\epsilon}_1^* \cr  
          \bar{\epsilon}-\bar{\epsilon}_1 &  \bar{\epsilon}-\bar{\epsilon}_1 &  \epsilon_2^2 \cr} 
 \right),
\end{eqnarray}
\noindent with 
\begin{equation}
\tilde{\epsilon}'=\frac{\epsilon_1^2\cos(2\alpha_1-\alpha_2)-
\epsilon_3^2\cos(2\alpha_3-\alpha_2)}{\epsilon_2}\ .
\end{equation}

Then, the squares of the neutrino mass eigenvalues  are  given as 
\begin{equation}
m_1^2\simeq \frac{h_0^4}{\epsilon_2^2 M_0^2}\ ,\qquad
m_2^2\simeq \frac{h_0^4}{\epsilon_2^2 M_0^2} 
                       (1-4\tilde{\epsilon}'+4\delta_1)\ ,\qquad
m_3^2\simeq \frac{h_0^4}{\epsilon_2^2 M_0^2} \epsilon_2^2\ ,
\end{equation}
\noindent
which are the spectrum called as  the inverted mass hierarchy.
Therefore, the ratio of the solar neutrino mass scale and 
the atmospheric neutrino mass scale is  given as
\begin{equation}
\frac{\Delta m^2_{\rm solar}}{\Delta m^2_{\rm atm}}=
\frac{m_2^2-m_1^2}{m_1^2-m_3^2}
\simeq 4(\delta_1  - \tilde\epsilon') \ .
\end{equation}
The experimental data of $\Delta m^2_{\rm solar}$ and $\Delta m^2_{\rm atm}$ 
\cite{Solar,Kamland,Atmospheric} give us an allowed region 
in the parameter space.

The unitary matrix, which diagonalizes the neutrino mass matrix of  
Eq. (\ref{mass}) such as  $V_\nu^T  M_\nu V_\nu = M_{\rm diagonal}$ , 
is given approximately by
 \begin{eqnarray}
 V_\nu\simeq 
 \left(
         \matrix{ 1 
                         & \frac{2\epsilon \cos{\varphi}+
\tilde{\epsilon}''}{2\delta_1} 
                         & \ord(\epsilon,\epsilon_i) \cr
                     -\frac{2\epsilon\cos{\varphi}+
  \tilde{\epsilon}''}{2\delta_1}
                         & 1 & \ord(\epsilon,\epsilon_i) \cr
             \ord(\epsilon,\epsilon_i) & \ord(\epsilon,\epsilon_i) & 1 \cr }
 \right),
 \end{eqnarray}
 \noindent with
 \begin{equation}
  \tilde{\epsilon}''=
 \frac{2\epsilon_1^2\cos{(2\alpha_1-\alpha_2)}-
  \epsilon_3^2\cos{(2\alpha_3-\alpha_2)}}{\epsilon_2}\ ,
   \label{def1}
 \end{equation}
\noindent where  $CP$ violating phases appear 
in (1,3), (3,1), (2,3) and (3,2) elements.

\subsection{Charged Lepton Mass Matrix}

When the three ${\bf 10}$'s and the Higgs multiplet $H_d$ distribute 
homogeneously in the bulk, one obtains the  democratic mass matrix 
for charged leptons.
Now we introduce distortions of  wave functions of the ${\bf 10}$'s 
in the bulk. We assume that dominant effects appear on the diagonal elements
of the above matrix. Then, the charged lepton mass matrix is given by 
\begin{equation}
M_\ell=\frac{m_0}{3} 
\left (\matrix{1& 1& 1 \cr 1& 1& 1\cr  1 & 1 & 1 \cr} \right )
+ \left (\matrix{\delta_{\ell 1}& 0&0 \cr 0& \delta_{\ell 2}& 0\cr
  0& 0  &\delta_{\ell 3}\cr} \right )  \ ,
\label{demo}
\end{equation}
\noindent where the second term of the right-hand side is the $S_3$ 
breaking  terms coming from the distortions of the ${\bf 10}$ fields. 
We take all $\delta_{\ell i}$ to be real, for simplicity.
This form of the mass matrix is used in Ref. \cite{Koide,FTY}. Here, we have 
assumed that the effects of distortion of the Higgs field $H_d$ are 
negligibly small.

The matrix of Eq. ({\ref{demo}) is diagonalized by $V_\ell=F_0 L_\ell$, where
\begin{eqnarray}
&& F_0= 
\left(\matrix{\frac{1}{\sqrt{2}} & \frac{1}{\sqrt{6}} & \frac{1}{\sqrt{3}}\cr
       -\frac{1}{\sqrt{2}} &  \frac{1}{\sqrt{6}} & \frac{1}{\sqrt{3}} \cr
               0       & -\frac{2}{\sqrt{6}} & \frac{1}{\sqrt{3}} \cr
                                         } \right )  \ , \nonumber \\
&& F_\ell \simeq   \left(
 \matrix{\cos\theta_\ell&\sin\theta_\ell&\lambda_\ell\sin 2\theta_\ell\cr
       -\sin\theta_\ell & \cos\theta_\ell &-\lambda_\ell\cos 2\theta_\ell \cr
 -\lambda_\ell\sin 3\theta_\ell&  \lambda_\ell\cos 3\theta_\ell & 1 \cr
                                         } \right )  \ , 
\end{eqnarray}
\noindent with
\begin{eqnarray}
&&\tan 2\theta_\ell \simeq \sqrt{3}
\frac{\delta_{\ell 2}-\delta_{\ell 1}}
{2\delta_{\ell 3}-\delta_{\ell 2}-\delta_{\ell 1}} \ ,
\qquad\qquad \lambda_\ell\simeq \frac{1}{3\sqrt{2}}\frac{\xi_\ell}{m_0}\ ,
\nonumber\\
&&\xi_\ell=\sqrt{(2\delta_{\ell 3}-\delta_{\ell 2}-\delta_{\ell 1})^2+
3 (\delta_{\ell 2}-\delta_{\ell 1})^2} \ .
\end{eqnarray}
The mass eigenvalues are given by
\begin{eqnarray}
&&m_e=\frac{1}{3}(\delta_{\ell 1}+\delta_{\ell 2}+\delta_{\ell 3})
-\frac{1}{6}\xi_\ell \ , \nonumber\\
&&m_\mu=\frac{1}{3}(\delta_{\ell 1}+\delta_{\ell 2}+\delta_{\ell 3})
+\frac{1}{6}\xi_\ell \ ,\nonumber\\
&&m_\tau=m_0+\frac{1}{3}(\delta_{\ell 1}+\delta_{\ell 2}+\delta_{\ell 3})\ .
\end{eqnarray}
\noindent 
If we take $\delta_{\ell 1}+\delta_{\ell 2}=0$ and 
$\delta_{\ell 2}\ll \delta_{\ell 3}$, for simplicity  \cite{FTY},
we get following mixings  in terms of the charged lepton masses:
\begin{equation}
\sin\theta_\ell \simeq -\sqrt{\frac{m_e}{m_\mu}} \ , \qquad
 \lambda_\ell\simeq \frac{1}{\sqrt{2}} \frac{m_\mu}{m_\tau}  \ ,
\label{mix0}
\end{equation}
\noindent which are  used in our numerical calculations.
It may be important to note that the condition
 $\delta_{\ell 1}+\delta_{\ell 2}=0$
 is crucial for the prediction of  $\sin\theta_\ell$.


\subsection{Right-handed Majorana Neutrino  Mass Matrix}

The right-handed Majorana neutrino mass matrix is examined
to discuss the leptogenesis.
Solving the hermitian matrix  $M_R M_R^{\dagger}$,
 \begin{eqnarray}
 &&M_R M_R^{\dagger}=M_0^2 \times \nonumber \\
\nonumber \\
 &&\left(
         \matrix{ \e_1^2+\e_2^2+\e_3^4 
         &  \e_1^2+2\e_2 \e_3^2 \cos{(\alpha_2-2\alpha_3)} 
         &\bar{\e}_1+\bar{\e}_1^* \bar{\e}_2+\bar{\e}_1^* \bar{\e}_3^2 \cr
                       \e_1^2+2\e_2 \e_3^2 \cos{(\alpha_2-2\alpha_3)}  
         & \e_1^2+\e_2^2+\e_3^4 
        & \bar{\e}_1+\bar{\e}_1^* \bar\e_2+\bar{\e}_1^* \bar{\e}_3^2 \cr
            \bar{\e}_1^*+\bar{\e}_1 \bar\e_2^*+\bar{\e}_1 \bar{\e}_3^{*2} 
        & \bar{\e}_1^*+\bar{\e}_1 \bar\e_2^*+\bar{\e}_1 \bar{\e}_3^{*2} 
         & 2\e_1^2+1 \cr }
 \right),
 \end{eqnarray}
\noindent we get
the squares of mass eigenvalues as follows:
 \begin{eqnarray}
 m_{R1}^2 &\simeq& M_0^2 [
                \e_2^2-4\e_1^2\e_2\cos{(2\alpha_1-\alpha_2)}+
 2\e_2 \e_3^2 \cos{(\alpha_2-2\alpha_3)}] \ ,\\
   m_{R2}^2  &\simeq & M_0^2
              [\e_2^2-2\e_2 \e_3^2 \cos{(\alpha_2-2\alpha_3)}]\ ,\\
 m_{R3}^2  &\simeq& M_0^2
[1+4\e_1^2\cos{2\alpha_1}+4\e_1^2 \e_2 \cos{(2\alpha_1-\alpha_2)}]\ .  
 \end{eqnarray}
 It is remarked  that the right-handed Majorana neutrinos of 
the first and second family are almost degenerated.
 The unitary  matrix to diagonalize  $M_R M_R^{\dagger}$  is given by 
 \begin{eqnarray}
 U_R\simeq
 \left(
 \matrix{ 
 \frac{1}{\sqrt{2}}& \frac{1}{\sqrt{2}} & \bar{\e}_1 \cr
 \frac{1}{\sqrt{2}}  & -\frac{1}{\sqrt{2}} & \bar{\e}_1  \cr
 -\sqrt{2}\bar{\e}_1^*  & 0 & 1 \cr}
 \right)  P  \ ,
 \end{eqnarray}
where  $P$ is the phase matrix, which is ignored in the following 
calculations because it does not affect our results.

 In order to discuss the leptogenesis later,
we take the diagonal basis of the right-handed Majorana neutrino mass matrix.
Then, the Dirac neutrino mass matrix $h_\nu$ is converted  to $U^T_R h_\nu$.
However, for our convenience, we use  $\overline h_\nu= U^T_R h_\nu U_R$
instead of $U^T_R h_\nu$ :
\begin{eqnarray}
{\bar h_{\nu}}
 &=& U_R^T h_{\nu} U_R\\
 &\simeq&  h_0
 \left(
 \matrix{ 
 1+\bar{\e} & -\frac{\bar\delta_1}{2} & 
 2\sqrt{2}i\bar\e_1\sin{\alpha_1}+\sqrt{2}\bar{\e} \cr
 -\frac{\bar\delta_1}{2} & 1-\bar{\e}+\frac{\bar\delta_1}{2} &  0 \cr 
2\sqrt{2}i\bar\e_1\sin{\alpha_1}+\sqrt{2}\bar{\e} & 0 & 1+\bar\delta_2 \cr}
 \right)   \ .       
\end{eqnarray}

\vskip 1 cm
\section{Numerical Study of Neutrino Mixings}

The lepton flavor  mixing matrix $U$ \cite{MNS} is obtained as
\begin{eqnarray}
 U = F_\ell^{\dagger} F_0^{\dagger} V_\nu \ ,
\end{eqnarray}
\vskip -0.5 cm
\noindent
which gives
\begin{eqnarray}
 && \sin\theta_{12}\simeq -\frac{1}{\sqrt{2}}+\sqrt{\frac{m_e}{m_\mu}}+
 \frac{1}{\sqrt{2}}\frac{2|\epsilon|\cos\phi+\tilde \epsilon}{2\delta_1} \ ,
\nonumber\\
&& \sin\theta_{13}\simeq \frac{2}{\sqrt{3}}\sqrt{\frac{m_e}{m_\mu}} \ , 
\nonumber\\
&&\sin\theta_{23}
\simeq -\frac{2}{\sqrt{3}}+\frac{1}{\sqrt{6}}\frac{m_\mu}{m_\tau} \ ,
\label{mix}
\end{eqnarray}
\noindent where
 $\theta_{ij}$ correspond to the mixing angles  in the conventional 
parameterization of the mixing matrix in  PDG \cite{PDG}.
 Putting the experimental data with   $3\sigma$   in~\cite{tortola,fogli},
\begin{eqnarray}
& & 7.2 \times 10^{-5}\, \mathrm{eV}^2
\leq  \Delta m_{12}^2 \leq  9.1 \times 10^{-5}\, \mathrm{eV}^2\ ,   \quad 
 0.23 \leq \sin^2{\theta_{12}} \leq 0.38\ , \nonumber \\
& &  1.4 \times 10^{-3}\, \mathrm{eV}^2
 \leq  \Delta m_{13}^2  \leq  3.3 \times 10^{-3}\, \mathrm{eV}^2\ ,  \quad
   \sin^2{2 \theta_{23}} \geq 0.90 \ , 
\label{data}
\end{eqnarray}
\noindent
 we obtain allowed regions of parameters in our model.

In the previous paper \cite{TY}, we  assumed that
 $\delta_1$ was  much larger than $\epsilon_i(i=1,2,3)$ and $\epsilon$.
Moreover, the only phase $\varphi$ was  taken account 
for the $CP$ violation in the analyses.
However, there is no reason to take  such assumptions.
Therefore, in this paper,
parameters $\delta_1$,  $\delta_2$,  $\epsilon_1$,  $\epsilon_2$,
 $\epsilon_3$ and $\epsilon$
are scanned in the region of $0$-$0.1$ 
 while phases $\alpha_1$,  $\alpha_2$,  $\alpha_3$,
$\beta_1$, $\beta_2$ and $\varphi$
are scanned in the region of $0$-$2\pi$.

Fig. 2 shows the plot of allowed region 
on the $\delta_1$-$\epsilon_i(i=1,2,3)$ and  $\delta_1$-$\epsilon$ planes.
\begin{figure}
\begin{center}
  \includegraphics[width=7 cm]{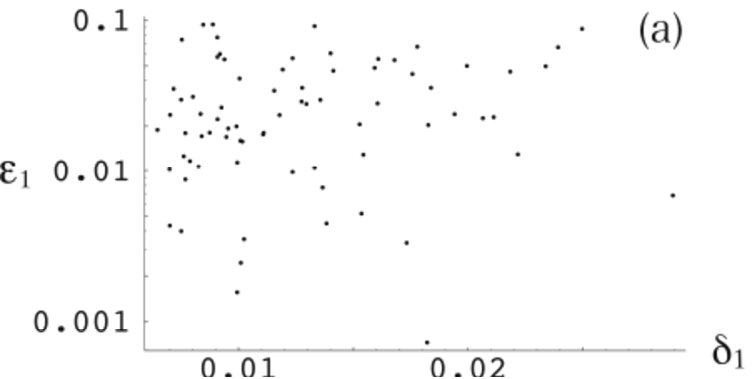} \hskip 1 cm
  \includegraphics[width=7 cm]{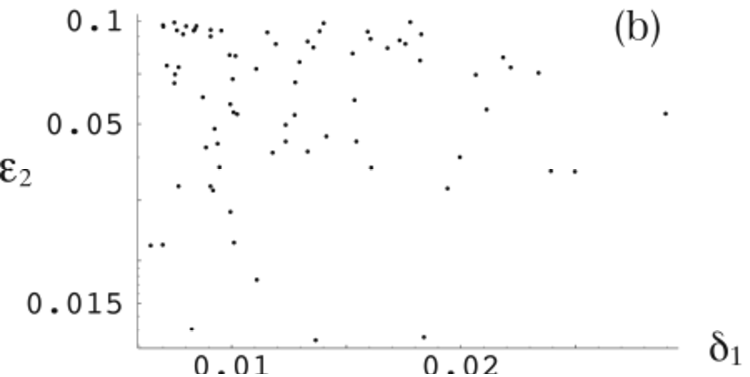} 
\vskip 1 cm
  \includegraphics[width=7 cm]{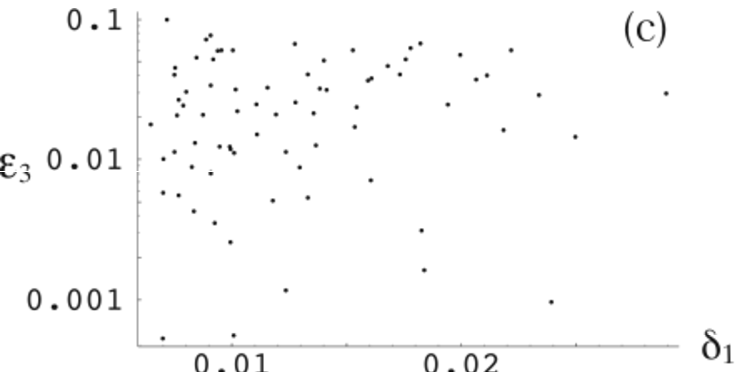} \hskip 1 cm
  \includegraphics[width=7 cm]{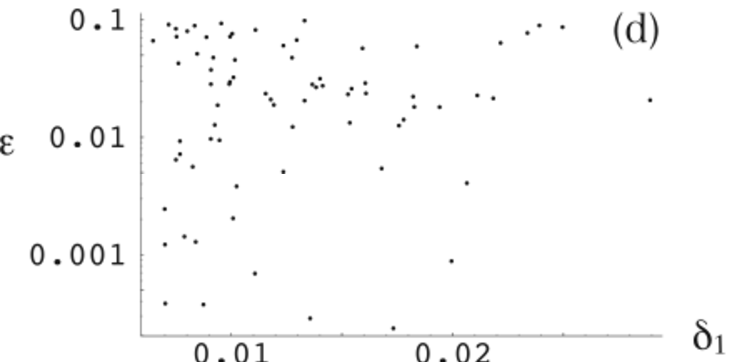}
\end{center}
\caption{Allowed regions on the  (a)  $\delta_1$-$\epsilon_1$ plane,
 (b) $\delta_1$-$\epsilon_2 $ plane, (c) $\delta_1$-$\epsilon_3$ plane
 and (d) $\delta_1$-$\epsilon$ plane.}
\end{figure}
The magnitude of $\delta_1$ is allowed in $0.003$-$0.025$ while 
$\epsilon_1$, $\epsilon_3$ and $\epsilon$ are allowed to be lower 
than $0.001$,  and  $\epsilon_2$ is larger than $0.01$. 
We omit figures of  allowed regions of phases 
since  all phases are  allowed in $0$-$2\pi$.

\begin{wrapfigure}{r}{8 cm}
\begin{center}
  \includegraphics[width= 7 cm]{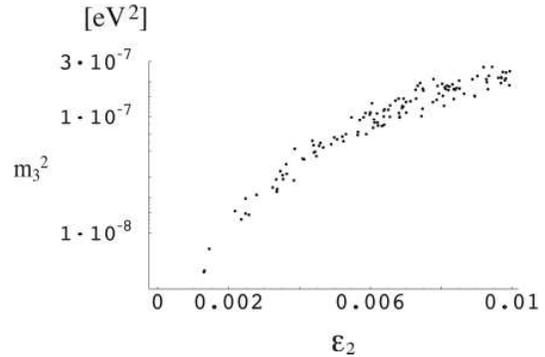} 
\end{center}
\caption{ The square of the lightest neutrino mass,  $m_3^2$ versus 
 $\epsilon_2$.}
\end{wrapfigure}

Since the model leads to the  inverted mass hierarchy for neutrinos,
the lightest neutrino mass is $m_3$, which depends mainly on 
 the parameter $\epsilon_2$.  Fig. 3 shows the $\epsilon_2$ 
dependence of $m_3^2$. The predicted $m_3$ is 
$ (1$-$ 50)\times 10^{-5} \ {\rm eV}$.


We can also discuss the neutrinoless double beta decay rate, 
which is determined by an effective Majorana mass:
\begin{equation}
\langle m \rangle_{ee}=\left|\ 
m_1c_{12}^2c_{13}^2e^{i \rho}+m_2s_{12}^2c_{13}^2e^{i\sigma}
+m_3s_{13}^2e^{-2i\delta_D} \ \right|\ ,
\end{equation}
\noindent
where $c_{ij}$ and  $s_{ij}$ denote
 $\cos \theta_{ij}$ and $\sin \theta_{ij}$, respectively, 
$\delta_D$ is a so called the Dirac phase, and $\rho,\sigma $ 
are the Majorana phases.
The Majorana phases are  estimated  from the mass matrix of Eq. (\ref{mass}).
Therefore, the Majorana phases are at most 
of the order of  $\epsilon_i(i=1,2,3)$ and  $\epsilon$, which are very small.
The predicted   $\langle m \rangle_{ee}$ is presented in Fig. 4,
where $m_1=0.05$ eV is fixed. This numerical  result is consistent with
the one  in the previous work \cite{TY}, 
$\langle m \rangle_{ee}\simeq  50 \ {\rm meV}$, which
 is  accessible to future experiments.

\begin{figure}
\begin{center}
 \includegraphics[width=7 cm]{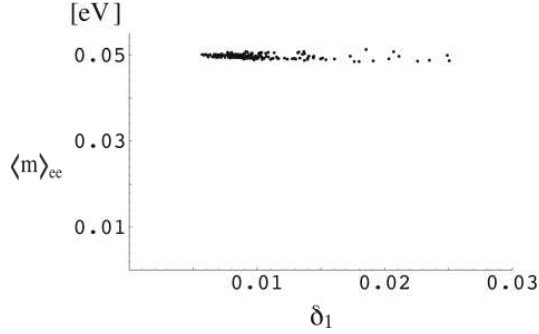} 
\end{center}
\caption{The predicted effective Majorana mass  $\langle m \rangle_{ee}$
versus $\delta_1$. }
\end{figure}

\section{Leptogenesis and $J_{CP}$}

In our scenario,  the  baryon asymmetry is explained by 
  the leptogenesis \cite{FY,gravitino} via decays of 
the right-handed neutrinos
$N_{i}$, which are produced non-thermally by decays of the inflaton
$\varphi_{inf}$ ~\cite{lep-inf}. 

Due to the non-vanishing phases in our model, CP invariance is violated 
in the Yukawa matrix $\bar h_\nu$. Then, the interference
between decay amplitudes of tree and one-loop diagrams results in the 
lepton number production \cite{FY}.
 The lepton number asymmetry per decay of the right-handed neutrino $N_{i}$
into the left-handed lepton doublets $l_{Lj}$ and the Higg scalar $H$ 
is given by \cite{FY,epsilon1}
\begin{eqnarray}
 \epsilon_{\ell i}
  &\equiv&
  \frac
  {\sum_j\Gamma (N_{i}\to l_{Lj} + H) 
  - 
  \sum_j\Gamma (N_{i}\to \overline l_{Lj} + \overline{H})}
  {\sum_j\Gamma (N_{i}\to l_{Lj} + H) 
  +
  \sum_j\Gamma (N_{i}\to \overline l_{Lj} + \overline{H})}
  \nonumber\\
 &=&
  -\frac{1}{8\pi}
  \frac{1}{( \overline h_\nu  \overline h_\nu^{\dagger})_{ii}}
  \sum_{k\ne i}
  {\rm Im}
  \left[
   \{
   \left(
    \overline h_\nu  \overline h_\nu^{\dagger}
    \right)_{ik}
    \}^2
   \right]
   \left[
    {\cal{F}}_{V}\left(\frac{m_{Rk}^2}{m_{Ri}^2}\right)
    +
    {\cal{F}}_{S}\left(\frac{m_{Rk}^2}{m_{Ri}^2}\right)
    \right]\,,
    \label{EQ-epsilon_i}
\end{eqnarray}
where  ${\cal{F}}_{V}(x)$ and ${\cal{F}}_{S}(x)$ represent contributions
from vertex and self-energy diagrams, respectively. 
In the case of the
SUSY theory, they are given by~\cite{epsilon1-SUSY}
\begin{eqnarray}
 {\cal{F}}_{V}(x) = \sqrt{x}\ln\left( 1 + \frac{1}{x}\right)
  \,,
  \qquad
  {\cal{F}}_{S}(x) = \frac{2\sqrt{x}}{x - 1}
  \, .
  \label{EQ-fs}
\end{eqnarray}
\noindent
Here, we  assume that the mass difference of the right-handed
neutrinos is large enough compared with their decay widths so that the
perturbative calculation is ensured
\footnote{In the quasi-degenerate case of the right-handed Majorana neutrinos,
 ${\cal{F}}_{S}(x)$ is modified as   
${\cal{F}}_{S}(x)=2(x-1)\sqrt{x}/\{(x - 1)^2+ (\Gamma^0_k)^2/m_{Ri}^2\}$ 
with $\Gamma^0_k\simeq h_0^2 m_{Rk}/(8\pi)$.
As far as $x-1>h_0^2$, this assumption is guaranteed. }.

The relevant elements of $h_{\nu}\bar  h_{\nu}^{\dagger}$   are given as 
\begin{eqnarray}
&&
 \rm{Im}[{(\bar h_{\nu} \bar h_{\nu}^{\dagger} )_{12}}^2]
= -\rm{Im}[{(\bar h_{\nu} \bar h_{\nu}^{\dagger} )_{21}}^2]
\simeq
2\e\delta_1^2 (\cos{\beta_1}+\frac{\delta_1}{4})\sin{(\varphi-\beta_1)} \ ,
\nonumber \\
&&(\overline h_\nu  \overline h_\nu^\dagger)_{11}
= |h_0|^2 (1+\delta_1-2|\epsilon| \cos\varphi) \ ,
\nonumber\\
&&(\overline  h_\nu  \overline h_\nu^\dagger)_{22}= 
|h_0|^2 (1+\delta_1+2|\epsilon| \cos\varphi)\ .
\label{hhdagger}
\end{eqnarray}
Since the right-handed neutrino masses of 
the first and second families  are almost degenerate  in the present model,
the self-energy contribution  ${\cal{F}}_{S}(x)$ is much larger than the 
vertex contribution $ {\cal{F}}_{V}(x)$ as follows:
\begin{eqnarray}
 {\cal{F}}_{V}(\frac{m_{R2}^2}{m_{R1}^2})&=&
 {\cal{F}}_{V}(\frac{m_{R1}^2}{m_{R2}^2})\simeq \ln 2 \ ,
 \\
 {\cal{F}}_{S}(\frac{m_{R2}^2}{m_{R1}^2})&=& 
-{\cal{F}}_{S}(\frac{m_{R1}^2}{m_{R2}^2})= 
      \frac{2m_{R1}m_{R2}}{m_{R2}^2-m_{R1}^2}\simeq \frac{1}{2}\
\frac{\epsilon_2}
{\epsilon_1^2 \cos{(2\alpha_1-\alpha_2)}-
\epsilon_3^2 \cos{(\alpha_2-2\alpha_3)}} \ .  \nonumber
\end{eqnarray}
\noindent
Therefore,  
the magnitude of ${\cal{F}}_{S}(\frac{m_{R2}^2}{m_{R1}^2})$ 
could be  enhanced by the quasi-degenerate right-handed neutrinos.
Then there occurs an enhancement of CP asymmetry for some region of 
the degeneracy. The scenario utilizing this enhancement is called 
as ``resonant leptogenesis'' \cite{reso1,reso2}.

\begin{figure}
\begin{center}
  \includegraphics[width=6.5 cm]{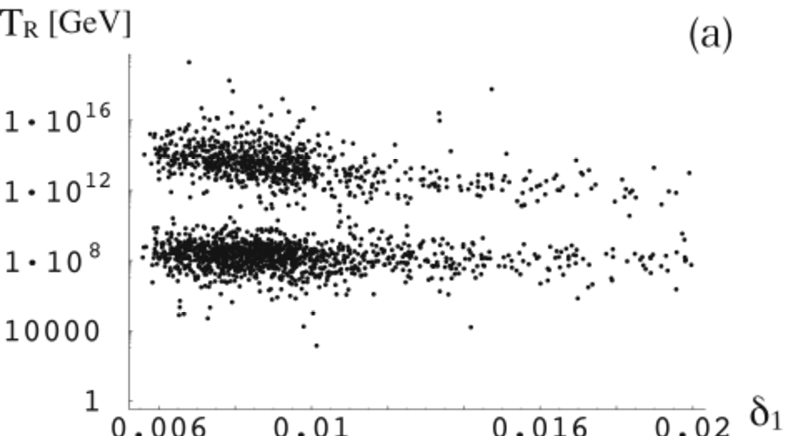} 
\hskip 1.5 cm
  \includegraphics[width=6.5 cm]{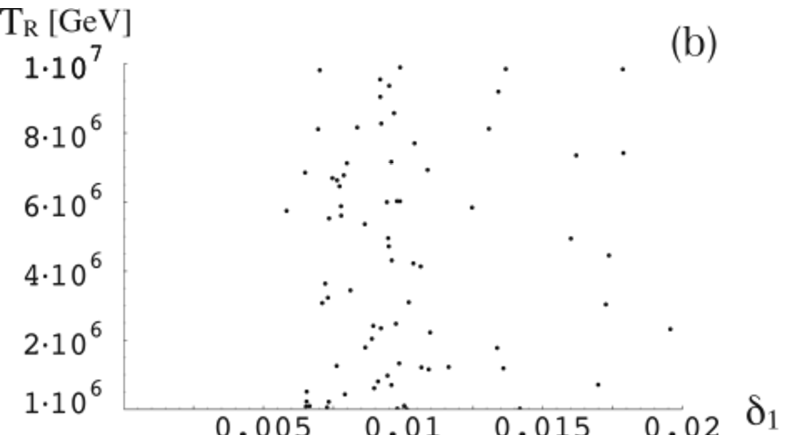} 
\end{center}
\caption{ Allowed  region of $T_R$ versus $\delta_1$ in order to explain
    the observed baryon asymmetry. The full region of $T_R$ is shown in (a),
while the $10^6$-$10^7$ GeV region of $T_R$ is shown in (b). }
\end{figure}

The lepton number asymmetry is given
\begin{eqnarray}
 \epsilon_{\ell 1}= \epsilon_{\ell 2}\simeq 
-\frac{|h_0|^2}{8\pi} \epsilon\delta_1^2
\left (\cos{\beta_1}+\frac{\delta_1}{4}\right )\ 
\frac{\epsilon_2}
{\epsilon_1^2 \cos{(2\alpha_1-\alpha_2)}-
\epsilon_3^2 \cos{(\alpha_2-2\alpha_3)}}
\sin(\varphi-\beta_1) .
\label{asym}
\end{eqnarray}
\noindent
The value of $|h_0|^2$ is given in terms of $m_1$ and $m_{R1}$:
\begin{eqnarray}
 |h_0|^2
\simeq  \frac{m_{1} m_{R1}}{(174\ \sin \beta \ {\rm GeV})^2}
 \ ,
\end{eqnarray}
\noindent
where  the vacuum expectation value  of the Higgs $H_u$ is taken as 
 $\langle H_u \rangle =174\sin\beta \  {\rm GeV} $
 ($\beta=\tan^{-1} \langle H_u \rangle/\langle H_d \rangle $).
The ratio of the lepton number density $n_L$ to the entropy density $s$
produced by the inflaton decay is given by~\cite{lep-inf}
\begin{eqnarray}
 \frac{n_L}{s} = \frac{3}{2}\sum_i
  \epsilon_{\ell i}
  B_r^{(i)}
  \frac{T_R}{m_{\phi_{inf}}}
  \,,
\label{lepto}
\end{eqnarray}
where $T_R$ is the reheating temperature after the inflation, $m_{\phi_{inf}}$
is the mass of the inflaton, and $B_r^{(i)}$ is the branching ratio of the
decay channel of the inflaton to $N_{i}$, i.e., $B_r^{(i)} =
B_r(\phi\to N_{i}N_{i})$. Here, we have assumed that the inflaton
decays into a pair of right-handed neutrinos, and $m_{Ri}>T_R$ in order
to make the generated lepton asymmetry not washed out by lepton-number
violating processes after the $N_{i}$'s decay.
 The dominant term in the right hand side of  Eq. (\ref{lepto})
 follows from $i=1$ and $2$ because $m_{R3}$ is much heavier than
  $m_{R1}$ and  $m_{R2}$. 
A part of the produced lepton asymmetry is immediately
converted~\cite{FY} into the baryon asymmetry via the ``sphaleron''
effect~\cite{sphaleron} since the decays of $N_{i}$ take place much
before the electroweak phase transition. The baryon asymmetry is given by
\begin{eqnarray}
 \frac{n_B}{s} = C\frac{n_L}{s}     \,,
\end{eqnarray}
where $C$ is given by $C \simeq - 0.35$ in the minimal SUSY standard
model ~\cite{LtoB}. Therefore,  $\epsilon_{\ell 1}$ should be negative 
in order to explain the sign of ${n_B}/{s}$.

 Now, we can obtain the allowed region of $T_R$  to reproduce 
the amount of the observed baryon asymmetry $n_B/s \simeq (0.8$--$0.9)\times
10^{-10}$~\cite{PDG} in our model.
Fig. 5(a) presents the allowed region of $T_R$ versus $\delta_1$,
where $2m_{R1(2)}/m_{\phi_{inf}}\simeq 1$ and $B_r^{(1)}+B_r^{(2)}\simeq 1$.
The  leptogenesis  works even in the  region  of $T_R=10^{4}$-$10^{6}$ GeV,
 which is lower than the expected
value,  $10^{7}$ GeV, in the previous paper \cite{TY}
 because of the  enhancement,  so called the resonant leptogenesis.
We also show the region of $T_R=10^{6}$-$10^{7}$ GeV  in Fig. 5(b).

\begin{wrapfigure}{r}{10 cm}
\begin{center}
  \includegraphics[width=8 cm]{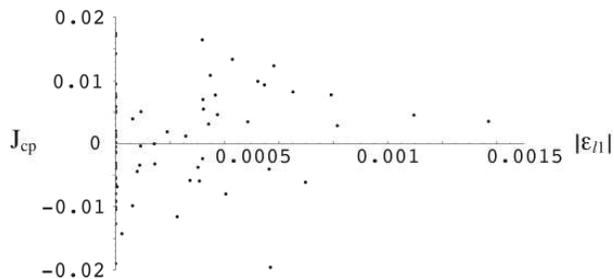}
\end{center}
\caption{Correlation between $J_{CP}$ and $|\epsilon_{\ell 1}|$.
   $\epsilon_{\ell 1}$ is taken to be
negative in order to give positive $n_B/s$. }
\end{wrapfigure}

 There is another $CP$ violation, which is the low energy phenomena
 and is expected to be measured in the future terrestrial experiments.
It is very interesting to ask whether there is the correlation between
observed baryon asymmetry and the leptonic $CP$ violation at the low energy.
In order to answer this question, 
 we plot  allowed region on the plane of  $\epsilon_{\ell 1}$ and 
 $J_{CP}$, which is the Jarlskog invariant \cite{Jcp} in Fig. 6.
We cannot find the  correlation between them as seen in Fig. 6.
Both positive and negative $J_{CP}$ are allowed for the negative 
$\epsilon_{\ell 1}$. 
The  $\epsilon_{\ell 1}$ is given mainly in terms of phases $\beta_1$ and
$\varphi$ as seen Eq. (\ref{asym})
 while  the magnitude of $J_{CP}$ is determined mainly by all phases 
except for $\varphi$.
For example, $|\epsilon_{\ell 1}|$ reaches $0.001$ but $J_{CP}$ is at most
$4\times 10^{-6}$ if $\varphi$ is the only non-vanishing phase.
On the other hand,  $|\epsilon_{\ell 1}|$ is at most  $10^{-8}$
 but $J_{CP}$ could be $0.01$ in the case of  non-vanishing
 $\alpha_1$,  $\alpha_2$,  $\alpha_3$ and $\beta_2$ with $\varphi=0$  and 
 $\beta_1=0$.
Only $\beta_1$ can contribute considerably 
  both  $\epsilon_{\ell 1}$ and  $J_{CP}$.


\section{Summary}

We have presented the comprehensive analyses of the model with
   inverted  hierarchy of the neutrino masses, which is based on the 
 ${\bf T^2/Z_3}$ orbifold model \cite{Geometry}. 
Here, a ${\bf 5}^*$ and a right-handed neutrino $N$  
in each family are localized on each of the equivalent three fixed points 
of the ${\bf T^2/Z_3}$ orbifold while three ${\bf 10}$'s and Higgs doublets 
$H_u$ and $H_d$ live in the bulk.  
The Higgs field $\phi$ responsible for the B-L breaking is assumed to be
  localized on the fixed point of the third family of ${\bf 5}^*$ and $N$.

While our analyses justify the qualitative result in the previous paper 
\cite{TY}, new results are also  obtained.
The $m_3$ is predicted in the region  $(1$-$50) \times 10^{-5}\ {\rm eV}$.
The model also  predicts  the element
of the neutrino mass matrix, $\langle m\rangle_{ee}$, responsible for 
neutrinoless double beta decays as $\langle m\rangle_{ee}\simeq 50$ meV,
which  has already been  predicted in the previous paper \cite{TY}.
We have shown that 
the observed  baryon asymmetry in the present universe is
produced by the non-thermal leptogenesis via the inflaton decay.
Due to the quasi-degenerate right-handed Majorana neutrino, 
 the baryon asymmetry is enhanced. 
In conclusion, we have found that the leptogenesis  works 
  even at the reheating temperature  $T_R=10^{4}$-$10^{6}\ {\rm GeV}$.
The low energy $CP$ violation $J_{CP}$ could be $0.02$,
however there is no correlation between $\epsilon_{\ell 1}$ and $J_{CP}$.

\paragraph{Acknowledgement}:

We thank T. Yanagida for suggesting numerical studies of the model.
We also thank H. Sawanaka for his help.
The work of M.T. has been  supported by the
Grant-in-Aid for Science Research
of the Ministry of Education, Science, and Culture of Japan
No. 17540243.


\end{document}